# Defect driven ferroelectricity and magnetism in nanocrystalline KTaO₃

I.S. Golovina[1]*, S.P. Kolesnik[1], V.Bryksa[1], V. Strelchuk[1], I.B. Yanchuk[1,2],

I.N.Geifman[3], S.A. Khainakov[4], S.V. Svechnikov[1], and A.N. Morozovska[1]

[1]*Institute of Semiconductor Physics of NAS of Ukraine, Pr. Nauky 41, 03028 Kyiv, Ukraine*

[2]*Nano Medical Technologies LLC, Gorkogo str. 68, 03150 Kyiv, Ukraine*

[3]*Quality Engineering Education, Inc., Buffalo Grove, IL 60089, USA*

[4]*Department of Organic and Inorganic Chemistry, University of Oviedo, 33006 Oviedo, Spain*

## Abstract

Nominally pure nanocrystalline KTaO₃ was thoroughly investigated by micro-Raman and magnetic resonance spectroscopic techniques. In all samples the defect driven ferroelectricity and magnetism are registered. Both ordering states are suggested to appear due to the iron atoms and oxygen vacancies. The concentration of defects was estimated to be 0.04 and 0.06-0.1 mole %, respectively. Note that undoped single crystals of KTaO₃ are nonmagnetic and have never exhibited ferromagnetic properties. The results enable us to refer a nanosized KTa(Fe)O₃ to the class of multiferroics and assume that it could perform the magnetoelectric effect at T<29 K. It was also established that the critical concentration of impurity defects necessary to provoke the appearance of the new phase states in the material strongly correlates with the size of the particle; as the size of the particle decreases, the critical concentration decreases as well.

# I. INTRODUCTION

The number of articles, including reviews, devoted to the research of multiferroics, i.e. materials, in which two or more order parameters coexist, drastically increased in recent years.[1,2,3,4] Primarily, these are materials, where such types of ordering as ferroelectricity, i.e. the spontaneous ordering of electric dipole moments, and (ferro)magnetism, i.e. the spontaneous ordering of spin magnetic moments, can coexist. From potential applications, especially attractive are the materials, in which room temperature ferromagnetism is observed.[5,6] Some compounds with perovskite structure also refer to multiferroics. Among others, potassium tantalate ($KTaO_3$) attracts a particular interest of researchers. It possesses a number of unique properties and therefore has been studied for a long time by different research groups, both experimental and theoretical.

Single-crystal $KTaO_3$ is an incipient ferroelectric that does not undergo a ferroelectric phase transition down to 0 K[7]. Although a crystal remains cubic in whole, there are regions, where cubic symmetry is broken even in nominally pure samples, so-called polar microscopic regions (or randomly oriented microdomains). Such polar microregions result in appearance of first-order Raman spectrum.[8] They also manifest themselves in such experimental data as the hysteresis of dielectric constant electric field dependence,[9] hysteresis in electric field effect in EPR of $Mn^{2+}$ in $KTaO_3$,[10] second harmonic generation (SHG)[11]. There are several suggestions which defects result in formation of the microdomains. These can be either intrinsic defects, like $Ta^{3+}$ or $Ta^{4+}$ and oxygen vacancy,[12,13] or impurity defects of type of niobium, iron or strontium, which are unavoidable impurities. Earlier,[14,15] even a ferroelectric phase transition at 10-13 K was observed in undoped crystals of potassium tantalate. Wemple[16] suggested that impurities are responsible for the phase transition. Commonly, doping of $KTaO_3$ with such elements as Li, Na, Nb, Sr, Ca etc., resulting in formation of solid solutions, substantially changing properties of this material, so that the incipient ferroelectric becomes a "real" ferroelectric with a wide variation of a Curie temperature. For example, in the system of $K_{1-x}Li_xTaO_3$ this range lies between 0 K for x=0 and 893 K for x=1. On the other hand, the appearance of ferroelectricity in undoped potassium tantalate becomes possible in the nanoscaled $KTaO_3$ crystal. As theoretical calculations show, a Curie temperature depends on the size and shape of nanoparticles.[17,18] Besides, in potassium tantalate doped with manganese, magnetoelectricity was discovered.[19] Features of its magnetic subsystem, being a spin glass, are similar to the $(Sr,Mn)TiO_3$ compound.[20] That allowed authors to attribute $K(Mn)TaO_3$ to the subclass of electro-magnetic materials, categorized disordered multiferroics. However, in undoped potassium tantalate any magnetic properties have not been noticed before.

Recently, Andriiko's group synthesized a nanocrystalline potassium tantalate with particles of 70-90 nm in diameter.[21] Ceramic samples were made from this powder, and some interesting effects were observed.[22] Firstly, a wide peak at 20-30 K was observed in the temperature dependence of dielectric constant. Maximum value of dielectric constant $\varepsilon_m$ was about 400 at 26 K. Since the dielectric peak did not shift to higher temperature as the measuring frequency was increased from 25 Hz to 1 MHz, it was suggested that the transition to the polar ordered phase takes place. Secondly, a ferromagnetic resonance (FMR) spectrum was registered on the same sample. Note that the peak-to-peak intensity of the spectrum did not increase with decreasing the temperature from 300 to 77 K. This result indicates the occurrence of a coupling between individual spins that leads to the long-range magnetic order. Also, some additional measurements were performed, which confirmed that some kind of magnetic order occurs. These results motivated us to investigate this material further.

In the present paper we report about the spectroscopic studies of nominally pure nanocrystalline potassium tantalate, Raman scattering investigation and the magnetic resonance (MR) study. From the first-order micro-Raman spectra the presence of microscopic ordered regions were found out. Its size increased markedly below $T_c=29$ K reaching up to 10-14 lattice constants at 7 K. The concentration of the polar defects, which are the source of the ferroelectric microregions in $KTaO_3$ nanopowder, are estimated to be 0.04-0.09 mole %, depending on the phonon parameter D~30-42 meV. Comparison analysis of the concentration between the Raman and the atomic emission spectrometry data allowed us to conclude that the softening $TO_1$ Raman mode are due to the symmetry-breaking Fe defects in the host lattice of $KTaO_3$. In the MR study two types of spectra, FMR spectrum originated from ferromagnetic centers and EPR spectrum originated from paramagnetic centers, are thoroughly investigated and interpreted. In the theoretical section, the dipole defect concentration necessary for the ferroelectric long-range order appearance was estimated using the random field theory, calculations by direct variational method were performed to solve the Schrödinger-Vanjie equation in the effective mass approximation. As a result, models of the defect driven ferroelectricity and magnetism in the nanosized $KTaO_3$ are suggested.

## II. MATERIAL AND TECHNIQUES

Nanocrystalline $KTaO_3$ was prepared by oxidation of tantalum metal powder (or $Ta_2O_5$ oxide) in potassium nitrate melt with additions of potassium hydroxide. The synthesis procedure is described in detail in Ref. [21]. After synthesis, the powder was additionally annealed at 1000 K in the air for 1.5 h to remove the residual moisture. X-ray diffraction (XRD) technique with $CuK_\alpha$ radiation, by using a DRON-3M diffractometer, confirmed that a perovskite-structured potassium metatantalate ($KTaO_3$) without any constituents of poly- or orthotantalates was formed. The sizes of the crystals were estimated from XRD data to be 70-90 nm.

Electron microscopic analysis was performed by the TEM method with a MET JEOL-2000 EX-II microscope operating at 200 kV. TEM investigation confirmed the XRD data. In Fig.1 the well-formed cubic crystals with dimensions not exceeding 100 nm are clearly seen.

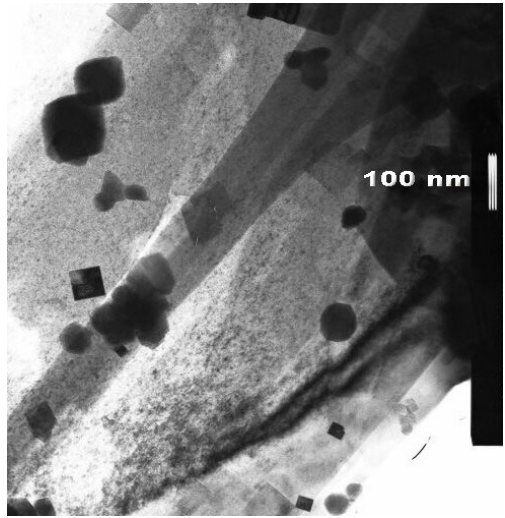

Fig.1. TEM image of the $KTaO_3$ powder

To produce ceramic samples, the powder was uniaxially pressed at 245 MPa into pellets of 5 mm in diameter and sintered at 1523 K for 2 h in the air. The obtained ceramics have a relative density about 80%.

Elemental analysis was performed with a Shimadzu ICPE-9000 inductively coupled plasma atomic emission spectrometer (ICP-AES). The operating conditions: Radio frequency 27.12 MHz; RF power 1.2 kW; quartz mini torch, plasma gas Ar 10,0L/min; auxiliary gas Ar 0.6 L/min; carrier gas Ar 0.7 L/min; view directions axial; exposure time 30 s. The powder of nanocrystalline $KTaO_3$ was placed in 100 ml of 0.1 mol/L $HNO_3$ for 1 h and after then it was filtered. High purity deionized water and high purity analytical reagents were used for preparations of the reference and sample solutions. The analysis showed the main unavoidable impurities (>0.0001 mole %) in the samples are Fe and Ni. The molar concentrations of Fe and Ni atoms were established to be no more than 0.04 and 0.008 %, respectively. Initial reagents and intermediate products of the $KTaO_3$ synthesis have also been analyzed, and the source of the impurities has been identified. These are tantalum metal powder, which was replaced with purified $Ta_2O_5$ oxide, and alumina crucible, in which the synthesis procedure was carried out.

Micro-Raman measurements were performed using Jobin-Yvon/Horiba *T*64000 Raman triple spectrometer equipped thermoelectrical-cooled charge-coupled device (CCD) detector. The 488.0 nm line of an $Ar^+/Kr^+$ laser was used as excitation source with power of 3 mW. An Olympus *BX*41 microscope equipped with a ×100 objective with $NA = 0.90$ at room temperature measurements and a ×50 long focus objective with $NA = 0.60$ at helium temperatures was used to focus the laser light on the sample and collect the scattered light into the spectrometer. We detected Raman shift smaller then $0.15\ cm^{-1}$ and were able to perform the Raman measurements in a low frequency range (>5 cm$^{-1}$). The temperature-dependent micro-Raman spectra were measured by using a CRYO Industries RC102-CFM cryostat.

Magnetic resonance experiments were performed on X-band CW EPR spectrometer Radiopan SE/X 2544 (Radiopan company, Poland) with a 100-kHz field modulation utilizing a standard rectangular $TE_{102}$ cavity. The magnetic resonance (MR) spectra of powder and ceramic samples were registered in the temperature range 20-620 K. The temperature interval of 20-80 K was achieved by blowing through the helium vapour in the cryostat and the temperature interval of 120-620 K was attained by blowing through the nitrogen vapour, controlled by the platinum resistance thermometers with accuracy of 0.5 K. Heat treatments of the powder samples at 873 K in the air for total of 12 h were made. In addition, samples were exposed to oxygen and helium atmospheres at 950 K. Temperature dependences of the MR spectra and the influence of the heat treatments are analysed below.

## III. EXPERIMENTAL RESULTS AND INTERPRETATION

### A. Paman scattering investigation

The KTaO$_3$ crystal has a cubic perovskite structure that belongs to the space group O$_h^1$ with the zero-wavevector phonons transform as $3F_{1u}+F_{2u}$ irreducible presentation.[23] As a result, first-order Raman scattering is forbidden in the electric-dipole approximation. In the Raman spectra of the powder samples of KTaO$_3$ display strong second-order Raman bands at 102.7; 252.6; 467.1; 575.4; 759.0 cm$^{-1}$ (Fig.2) which complicates an observation of the phonons. However, these bands narrowed at helium temperatures and we detected the longitudinal and transverse LO$_1$, TO$_2$, LO$_2$, TO$_3$, LO$_3$, TO$_4$, LO$_4$ phonon modes at ~116.1; 197.7, 275.9; 299.1; 485.6, 547.4; 813.3 cm$^{-1}$, respectively (Fig.2). Such phonons were observed usually for KTaO$_3$ single crystals in an external electric field applied in the [001] direction [23]. The electrical field distorted the KTaO$_3$ unit cell and caused the transformation of the point group of the crystal to the tetragonal group C$_{4v}^1$. Observed phonon modes in the Raman spectra on the Fig.2 are due to the symmetry-breaking defects which induce ferroelectric microscopic regions in the powder samples of KTaO$_3$ [8, 24].

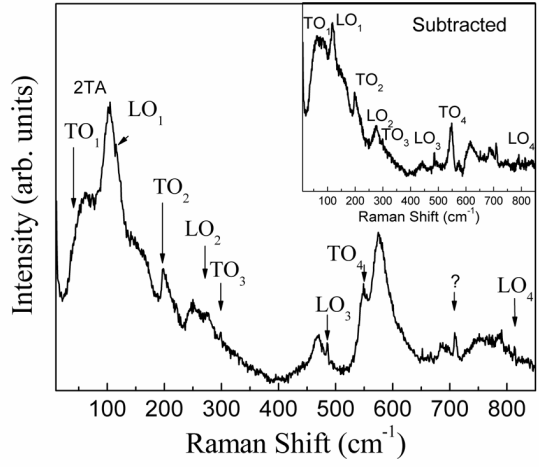

Fig.2. The micro-Raman spectra of the powder samples of KTaO$_3$ measured at T=7 K. Inset illustrates the subtraction the second-order Raman bands from the spectra. The phonon peak at 708.7 cm$^{-1}$ is not determined (denoted by ?). $\lambda_{exc}$ = 488.0 nm. P$_{exc}$ = 3 mW.

It is known that a dielectric constant $\varepsilon_0$ of the ferroelectric crystals is strongly temperature-dependent function and vanishes in a ferroelectric phase transition point as $1/\varepsilon_0 = (T - T_c)/C$ [25]. At that the lowest transverse TO$_1$ phonon becomes a soft-mode with a frequency $\omega_{soft}(TO_1)$ and we can write Lyddane-Sachs-Teller equation for the case several optical modes in the ferroelectric crystals as follows [23, 25]:

$$\omega_{soft}^2(TO_1) = \frac{\varepsilon_\infty}{\varepsilon_0}\left[\prod_i (\omega_{LO})_i^2 / \prod_i (\omega_{TO})_{i\neq 1}^2\right] \approx A(T - T_c) \,. \tag{1}$$

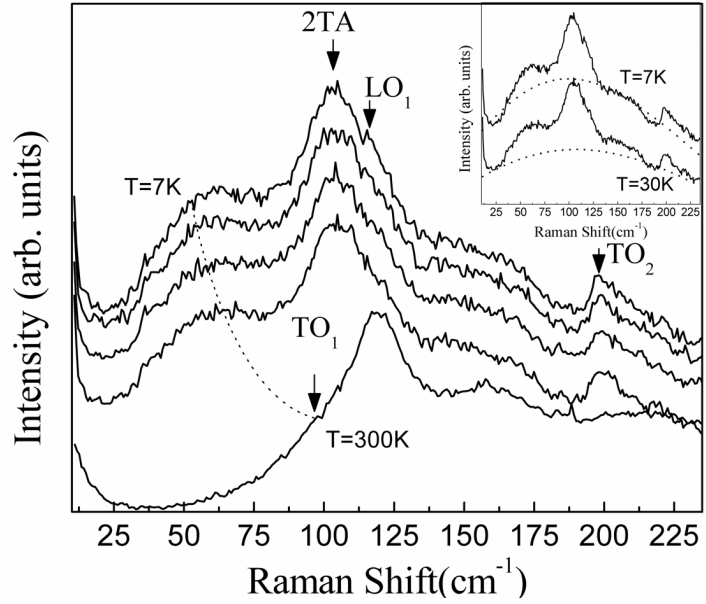

Fig. 3. Micro-Raman spectra of the powder samples of $KTaO_3$ measured at T = 7; 15; 25; 30; 300 K (from top to bottom). Inset shows the comparison of calculated and experimental spectra for various temperatures. For the analysis of the $TO_1$ phonon mode, the second-order Raman band due to 2TA phonons was subtracted at the modulation of the $TO_1$ soft-mode.

To study the softening of the $TO_1$ phonon mode, we performed Raman measurements in the temperature interval 7<T<300 K (Fig. 3). Under the temperature decrease we detected the softening of the $TO_1$ phonon at temperature less than $T_c \sim 29K$ (as will be found below). The disorder induced Raman spectra shape and intensity depend on the magnitude and spatial extent of the static distortion induced by the symmetry-breaking defects [8]. Due to this fact, we are able to carry out a quantitative analysis of the defects, and find their concentration in the host lattice of $KTaO_3$.

As shown in Ref. [8] the scattered spectrum of the first-order Raman processes for the $KTaO_3$ is caused by the static distortion around the symmetry-breaking defects and the Raman intensity can well be described in terms of the defect-induced correlation function $S(q)$ and phonon density of states $\left[\left(d\omega_q/dq\right)_{q=q_\omega}\right]^{-1}$ as follows:

$$I(\omega) = \frac{(1-e^{-\omega/T})^{-1}}{\omega} q_\omega^2 S(q_\omega) \left[\left(\frac{d\omega_q}{dq}\right)_{q=q_\omega}\right]^{-1},$$  (2)

where $\omega_q$ is a dispersion curve of the lowest $TO_1$ frequency of $KTaO_3$, which can be written as $\omega_q^2 = A(T-T_c) + Dq^2$ [22]. D is a phonon parameter that was regarded as an isotropic and temperature independent in our fitting procedure. $T_c$ is a temperature of the ferroelectric phase transition. A is a Curie-Wiess constant (see Eq. 1).

We have chosen the simplest form for the defect-induced correlation function $S(q)$ to provide an adequate fitting of the Raman spectra [8]:

$$S(q) = \frac{S_0}{(1+q^2R_0^2)^2}$$  (3)

This expression for $S(q)$ results from a random field approach which holds only as long as the correlation range of the random defect potential field itself is small compared to size $R_0$ of microscopic ferroelectric regions ($q > 1/R_0$).

Combining together Eqs. (2) and (3) we obtain the final expression for the Raman spectra fitting (see inset in Fig. 3)

$$I(\omega) \sim (1-e^{-\omega/T})^{-1} \frac{(2\pi)^2}{R_0 a^2} \frac{Q}{(R_0^{-2}+Q^2)^2},$$  (4)

where $Q = (\omega^2 - A|T-T_c|)/D^2$ and $a$ is a lattice constant of $KTaO_3$ ($a$=3.988 Å [22]).

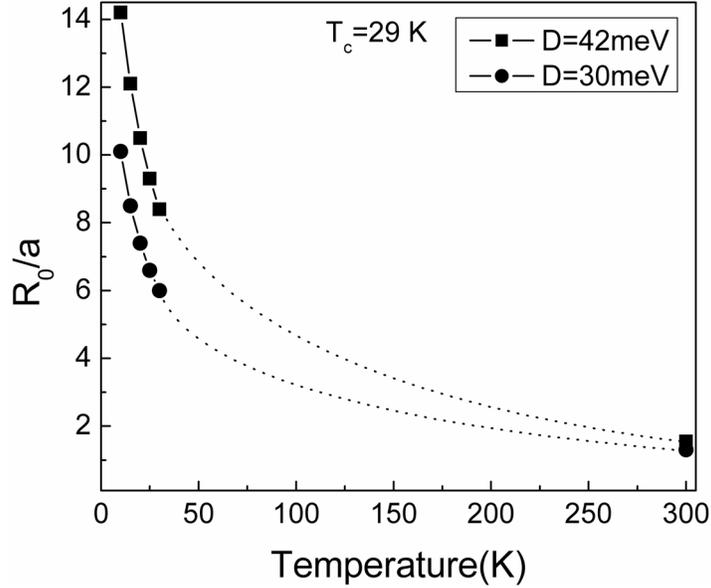

Fig.4. Size of the microscopic ferroelectric regions as a function of temperature, determined by the fit of Eq.(4) to the experimental spectra, as the inset in Fig. 3 using the phonon parameters D=42, 30 meV. Obtained value of the ferroelectric temperature point in the fits is $T_c$=29 K.

Fig. 4 shows the obtained values of $R_0/a$ and $T_c$ versus temperature at fixed D and A parameters in Eq.(4), which were determined from independent experiments [23, 8]. The ferroelectric region size increases dramatically as the temperature is lower than the ferroelectric transition point (T<$T_c$). The size $R_0$ of the microscopic ferroelectric regions in $KTaO_3$ nanopowder is changed from 10 to 14 lattice constants with increases of the phonon parameter D from 30 to 42 meV. This corresponds to the variation of the molar concentration of the symmetry-breaking defects from 0.09 to 0.04% (Fig. 3). By changing the phonon parameter D of modulation of the $TO_1$ soft mode in the Raman spectra we take into account anisotropy effects, which improve the accuracy for the defect concentration determination. It is important to say that the concentration value of the defects obtained by fitting of the $TO_1$ softening mode (Fig. 4) agrees well with the elemental analysis provided by ICP-AES, which shows a molar concentration of the Fe impurities about ~0.04% in the $KTaO_3$ powder sample. This allows us to suppose that the impurities, which induce the softening Raman mode, are the Fe atoms in the $KTaO_3$ nanopowder.

## B. A magnetic resonance study

As it was defined in our previous paper [22], the magnetic resonance spectrum of nanocrystalline potassium tantalate consists of two types of signals, a broad ferromagnetic resonance (FMR) signal, and an electron paramagnetic resonance (EPR) signal. Below the nature of both signals are thoroughly investigated. The spectra were registered on a powder sample in the temperature interval 20<T<620 K (Fig.5) and on a ceramic sample at T=77 and 300 K (Fig.6). First, let us consider the behaviour of the FMR spectrum. As shown in Fig.5, a FMR spectrum is a broad temperature dependant resonance signal. The spectrum is isotropic in powder sample with g=2.23 and peak-to-peak linewidth ΔB=1000 G at room temperature. Note, that the resonance field $B_{res}$ was defined as the apparent intersection of the absorption derivative with the base line. The arrow with asterisk in Fig.5a denotes a paramagnetic impurity signal at g≈4.3 originating from the quartz tube (probably due to $Fe^{3+}$ ions) filled with a powder sample. Because of the known behaviour of this signal, it was used as a reference paramagnetic signal. As seen from Fig.5, the FMR spectrum broadens and shifts to lower fields with the temperature decrease. It becomes rather broad with g=2.31 and $ΔB_{pp}$=1500 G at T=20 K. Temperature dependences of g-factor value and peak-to-peak linewidth ΔB are presented in Fig.6. Both parameters monotonically increase with decreasing the temperature from 620 to 120 K and become constant below 120 K. The integral intensity of the FMR spectrum remains constant over all interval of 20-620 K. Note that integral intensity of the common EPR signal, which is proportional to the magnetic susceptibility χ of the paramagnetic center, increases according to a Curie law ($\chi \sim 1/T$). The FMR signal exhibits evident shape anisotropy in ceramic sample . As shown in Fig.7a, it resonates at g=2.35 when static magnetic field $B_0$ is perpendicular to pressure direction, P, (line 1) and at g=2.08 when $B_0$ is parallel to P (line 2). Similar to the powder sample, in the sintered sample the FMR signal shifts to lower fields as the temperature decreases. As shown in Fig.7b, at T=77 K it resonates at g=2.81 when $B_0 \perp P$ and at g=2.38 when $B_0 \| P$.

Powder

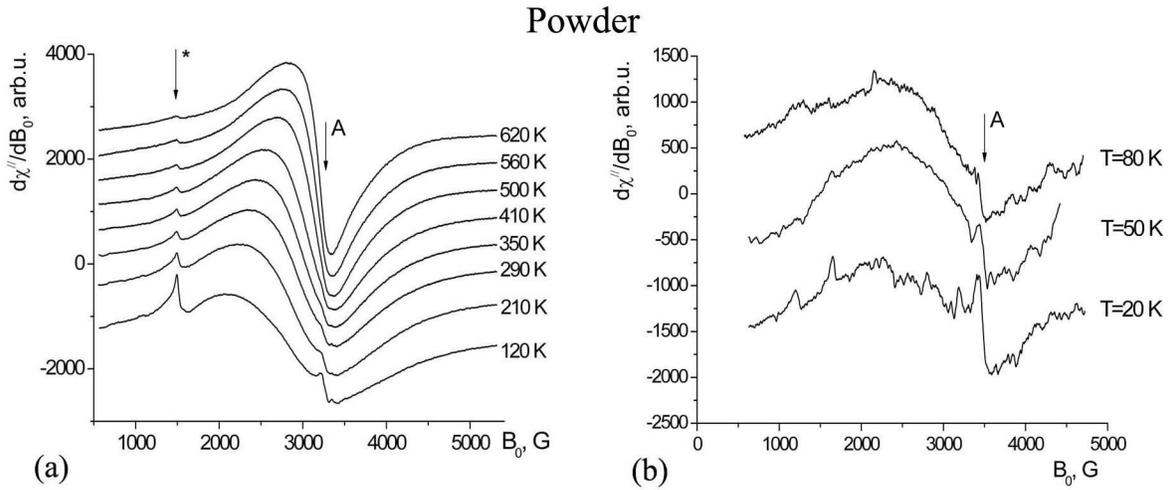

Fig.5. Magnetic resonance spectra of KTaO$_3$ powder sample recorded in the temperature intervals of (a) 120-620 K at $\nu$=9101 MHz and (b) 20-80 K at $\nu$=9611 MHz.

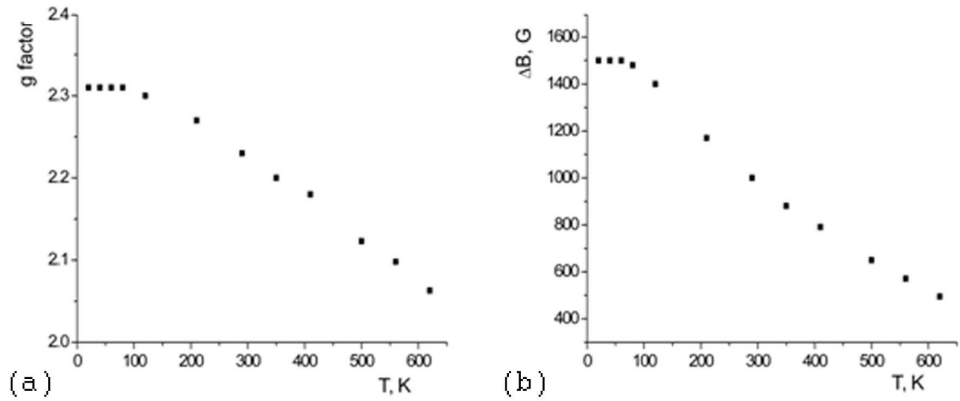

Fig.6. Temperature dependences of (a) g-factor value and (b) the linewidth ΔB of the FMR

spectrum recorded on the powder sample.

## Ceramics

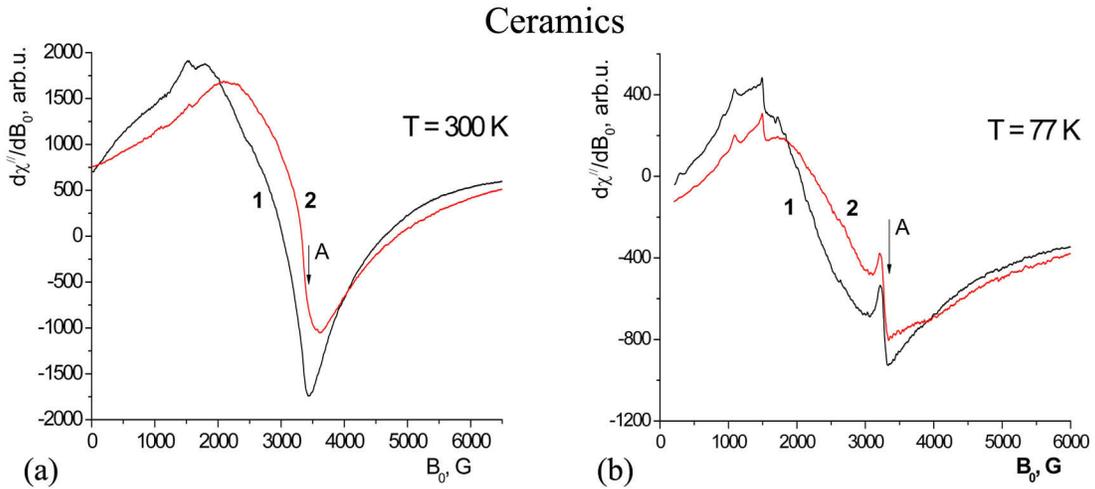

Fig.7. (Color online) Magnetic resonance spectra of KTaO$_3$ ceramic sample registered at: (a) T=300 K, $\nu$=9385 MHz and (b) T=77 K, $\nu$=9122 MHz in two orientations of static magnetic field (B$_0$): B$_0$⊥P (line 1, black) and B$_0$∥P (line 2, red). P is pressure direction.

Before the interpretation of our results, let us describe the peculiarities of the FMR in nanoscaled systems. It is known that temperature behavior of the FMR signals has some new properties in nanosized materials. Actually, it represents a superparamagnetic resonance (SPR), even though it still denotes as a FMR in many articles.[26,27,28,29] We will briefly describe the features of the SPR (or a FMR of superparamagnetic nanoparticles). Comprehensive theoretical developments on this problem were made by Raikher et al.[30] and Biasi, et al.[31]. Berger et al.[32,33] elaborated the approach for the computer-assisted analysis of magnetic resonance spectra of superparamagnetic materials. As well known, the main characteristic of a ferromagnetic material is a spontaneous magnetization, i.e. the existance of magnetization in the absence of external magnetic field (if the material is magnetized to saturation, i.e. being in the monodomain state). In nanosized particle which volume is less than $V=10^{-18}$ cm$^3$, the domain structure is not formed,[34], e.g. such a particle is uniformly magnetized, and its magnetic moment is $\mu=M_sV$, where $M_s$ is the magnetization of the material. Typically, the value of $\mu$ amounts from several hundreds to several thousands of Bohr magnetons $\mu_B$, and it is called a superspin. The direction of a superspin is defined by the resultant field, which includes the anisotropy and demagnetizing fields of the particle, as well as the coupling with surrounding. At low temperatures the magnetic moment of the particle is "frozen" (so called "blocking" effect), and the directions of the superspins are randomly distributed. In this state the overall magnetization of the material drops almost to zero. At this point, the SPR line of the nanoparticles assembly strongly broadens and then becomes invisible as temperature decreases. When temperature increases thermal fluctuations of the direction of a superspin decrease the role of the anisotropy due to the angle averaging, and at the temperatures $T>T_B$ ($T_B$ is a "blocking" temperature) the SPR line narrows. Thus, the main features of the behavior of the SPR signal, that distinguish it from the FMR signal, are the decrease of the integral intensity $I$ ($I \sim \chi$) as the temperature decreases and a wide broadening and consequent disappearing of the signal below $T_B$. Other peculiarities of the SPR signals namely the resonance field shifting and the linewidth narrowing with increasing the temperature are also specific for the FMR signals.

Based on the brief analysis of the superparamagnetic state, we can declare, that not a superparamagnetic, but a ferromagnetic spin ordering takes place in our system. In support of this statement there are two results. First, an integral intensity of the FMR spectrum remains constant between 620 and 20 K. Second, the "blocking" effect does not occur. Further, the presence of the stable magnetization in the absence of external magnetic field at room temperature was revealed from the following experiment.

The powder of $KTaO_3$ was mixed with paraffin and placed into the quartz tube. The initial spectrum of the mixture recorded at T=300 K was isotropic. Then the mixture was heated up to T=380 K. After that, the tube containing the mixture was magnetized using a static magnetic field of about 0.65 T and cooled down to T=300 K. Then the spectrum was recorded again, and the angular dependence was observed. Magnetic anisotropy was directed along the polarizing field. Subsequent heating and cooling of the tube otherwise orientated with respect to the magnetic field result in a relevant change of the anisotropy direction of the FMR signal. Thus, the experiment and above data together confirmed the presence of long-range magnetic order in the powder.

It should also be noticed about a spin glass as a possible state. If so, then a magnetic susceptibility of the material would increase with the temperature decrease. However, it did not happen in our experiments.

To determine the nature of the FMR signal, the heat treatments of powder samples in the air and in oxygen and helium atmospheres were made.

First, the powder was heated in the air at 870 K. As Fig.8 shows, the integral intensity $I$ of the FMR spectrum decreases at about 14% after the heating for total of 12 h.

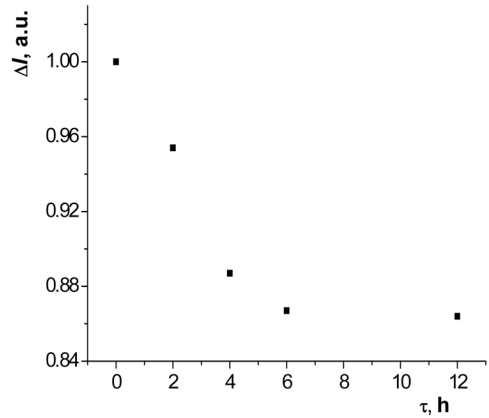

Fig.8. Change of the integral intensity of the FMR spectrum versus the heating time in air.

Then, the heating in pure oxygen was made. The integral intensity of the FMR spectrum decreased by two times (about 50%) after heating at 950 K for 3.5 h. As the effect of the heating in oxygen is about 5 times higher than in air, that proved the direct influence of oxygen. Since the intensity remains almost constant after 4 h of heating in air (see Fig.8), we presume that heating in the oxygen atmosphere for 3.5 h is enough to achieve the effect. In Fig.9 the spectra, before (line 1) and after (line 2) the heating in oxygen, are presented. It should be noted that the FMR spectrum has been totally restored after subsequent storing the sample in air at room temperature. Further heat treatments in the helium atmosphere and in vacuum did not influence the spectrum. Decreasing the intensity of the FMR spectrum after the heating in the oxygen atmosphere reveals the direct participation of the oxygen vacancies in the ferromagnetic ordering and/or in the structure of the ferromagnetic center. However, the ferromagnetic center also contains the magnetic ions, most likely $Fe^{3+}$, concentration of which remains constant. That is why the helium and vacuum did not influence the spectrum. We can also say that the ferromagnetic centers are located partly in the vicinity of the particle surface and partly in the deeper layers. Firstly, it is because the spectrum is totally restored after storing the sample in air at room temperature. Secondly, because the integral intensity of the spectrum, and hence the quantity of the ferromagnetic centers, decreases only at 50%, not at 100%. In fact, more rigorous conditions of the heating (temperature, partial pressure of oxygen and the duration) would destroy the spectrum completely.

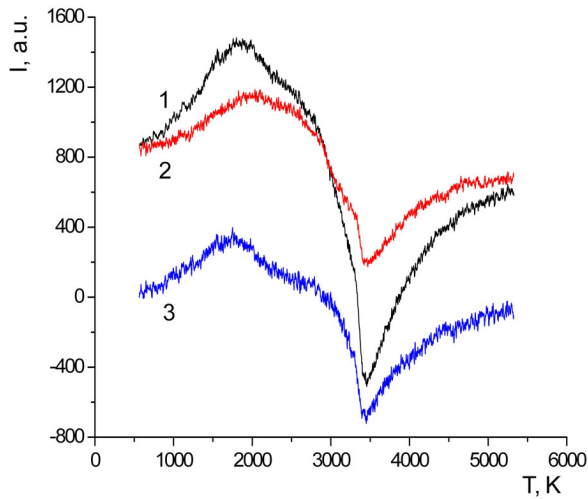

Fig.9. (Color online) FMR spectra of $KTaO_3$ powder sample before (spectrum 1) and after (spectrum 2) heating in oxygen atmosphere. Spectrum 3 is the difference of the spectra 1 and 2.

Now let us interpret the EPR spectrum for powder and ceramic samples. Arrow with letter A in Figs.5 and 7 indicates the position of the spectrum at g=1.996±0.001. Its peak-to-peak linewidth is 90 G. The resonance field is temperature independent, and the integral intensity follows a Curie law. Evidently, the paramagnetic centers are the source of the signal. The concentration of the paramagnetic centers exhibited the EPR spectrum in $KTaO_3$ nanopowder is 0.003 mole %. We focus at two possible origins of the paramagnetic centers, namely Fe and Ni ions as unavoidable impurities in our samples. Note that Fe is the most common uncontrolled paramagnetic impurity, observed by the EPR method in $KTaO_3$ single crystals. Usually these are individual paramagnetic ions of $Fe^{3+}$ and $Fe^+$ incorporated into the host lattice. The $Fe^{3+}$ ions form the EPR centers of different symmetry, from cubic to rhombic, and substitute both $Ta^{5+}$ and $K^+$.[35,36,37,38,39,40] Also, in nominally pure single crystals of $K_{1-x}Li_xTaO_3$ the EPR spectrum of $Ni^{3+}$ was observed.[41] Earlier, Hanon[42] studied the EPR of $Ni^{3+}$ in doped single-crystal $KTaO_3$ and defined that $Ni^{3+}$ ions at the $Ta^{5+}$ site are in low-spin configuration and undergoes a Jahn-Teller axial distortion. We will consider low-symmetry axial and rhombic centers of $Fe^{3+}$ and axial centers of Fe+ and $Ni^{3+}$. The EPR parameters of these centers in $KTaO_3$ single crystals are presented in the Table 1.

Table 1. Parameters of EPR centers of $Fe^{3+}$, $Fe^+$ and $Ni^{3+}$ in $KTaO_3$

| No. | Model of center | Symmetry | Spin-Hamiltonian parameters | Ref. |
|---|---|---|---|---|
| 1 | $Fe^{3+}(Ta) - V(O)$ $\delta z = 0.15$ Å | axial | $g_\perp$=1.996(2), $g_{||}$=2.00(1) $b^0_2$=D=1.33 cm$^{-1}$ (300 K), $b^0_2$=D=1.44 cm$^{-1}$ (77 K) | [35] |
| 2 | $Fe^{3+}(K) - O^{2-}$(int.) | axial | $g_\perp$=2.004(2), $g_{||}$=2.00(1) D=4.46 cm$^{-1}$ | [36] |
| 3 | $Fe^+$ (K) (off-center) | axial | $g_\perp$=2.16, $g_{||}$=2.02 (S=3/2) $g_{\perp eff}$=4.33, $g_{||eff}$=2.02 ($S_{eff}$=1/2) | [38] |
| 4 | $Fe^{3+}(Ta) - 2V(O)$ $\delta z = 0.25$ Å | rhombic | $g_x$=1.99, $g_y$=2.01, $g_z$=2.00, $b^0_2$=D=0.44 cm$^{-1}$, $b^2_2$=3E=0.196 cm$^{-1}$, $b^0_4$=1.9*10$^{-4}$ cm$^{-1}$, $b^2_4$=1.7*10$^{-2}$ cm$^{-1}$, $b^4_4$=1.8*10$^{-2}$ cm$^{-1}$ | [39] |
| 5 | $Fe^{3+}(K) - V(O)$ in [110] direction | rhombic | $g_{eff}$=4.27 | [40] |
| 6 | $Ni^{3+}$ | axial | $g_{\perp eff}$=2.111, $g_{||eff}$=2.234 | [42] |
| 7 | $Ni^{3+}$ | axial | $g_{\perp eff}$=2.086, $g_{||eff}$=2.169 | [42] |

$\delta z$ denotes a displacement of $Fe^{3+}$ ion from the center of $Ta^{5+}$ site

We simulated the powder EPR spectrum with each set of parameters, and found that the centers #1, 2 and 4 presented in Table 1 are most likely the sources of the EPR spectrum observed, because they have the g-factor values, which are closest to g=1.996. Note that the EPR spectra of the centers #1 ($Fe^{3+}$ substitutes for $Ta^{5+}$ with nearest oxygen vacancy V(O)) and 2 ($Fe^{3+}$ substitutes for $K^{+}$ with interstitial $O^{2-}$ ion) coincide in the X-band EPR experiments [35]. Therefore, it is unsurprisingly that they coincide in the powder EPR spectrum also. The EPR center of rhombic symmetry has very close parameters (especially to the center #1), so its powder spectrum is located almost at the same g-factor value. Come back to the MR spectra before and after heating in oxygen, presented in Fig.9. Comparison the shape of line 1 and line 2 in Fig.9 shows us that the intensity of the EPR spectrum decreases in 2 times after the heating and restores after staying in air at room temperature, similar to the FMR spectrum. Such behavior evidence that the EPR center contains an oxygen vacancy. Among the centers #1, 2 and 4, only the centers #2 and 4 have the oxygen vacancy. Easy restore of the spectrum after the heating indicates that these are surface centers. However, part of the centers is located in the bulk, since the spectrum intensity decreases at about 50% after the heating in the oxygen atmosphere.

### IV. THEORETICAL RESULTS

#### A. Ferroelectric long-range order in KTaO₃

Appearance of ferroelectricity in the incipient ferroelectrics like $KTaO_3$ with unavoidable dipole impurities was well explained by the random field theory [43]. Actually, the randomly situated defects (impurity atoms, dilatation centers, etc) act as randomly situated dipoles, which start to interact via the soft phonon mode to state the long-range ferroelectric order under the favorable conditions (dipole concentration more than critical for the fixed temperature, etc).

Using the random field theory one could estimate the ferroelectric phase transition temperature $T_c$ in the incipient ferroelectrics with known defect concentration $n$, namely [43]:

$$k_B T_c \approx \frac{n}{4\pi\varepsilon_0\varepsilon}\left(\delta z\, Z\, e\left(1+\frac{\varepsilon-1}{3}\right)\right)^2 \approx \frac{n\varepsilon}{4\pi\varepsilon_0}\left(\frac{\delta z\, Z\, e}{3}\right)^2 \qquad (5)$$

Where $\delta z$ is the dipole length, $\varepsilon(T_c)$ is the ferroelectric host lattice dielectric permittivity at temperature $T=T_c$, $Z$ is the defect charge, $e$ is the elementary charge, $\varepsilon_0$ is the universal dielectric constant. Calculations typically give values $T_c = (10\text{-}30)$ K at $n \leq 1\%$. Using the expression (1) for temperature $T_c$, it is easy to get the defect concentration as $n \approx 4\pi\varepsilon_0 \dfrac{9 k_B T_c}{\varepsilon(\delta z\, e)^2}$.

Using the random field theory let us estimate the dipole defect concentration $n$ necessary for the ferroelectric long-range order appearance at temperature $T_c = 29$ K, that corresponds to the maximum of KTaO$_3$ nano-ceramics dielectric permittivity $\varepsilon(T_c) = 400$ obtained in our previous experiments [22]. Estimations are made for two defect centers (see Table 1), which are considered below as possible origin of the polar nanoregions.

(a) For the defect center **$Fe^{3+}(Ta) - V(O)$ (axial)** the parameters are: $T_c = 29$ K, $\varepsilon(T_c) = 400$, $\delta z = 0.15$Å$=1.5\ 10^{-11}$ m, $Z=3$ elementary charges. For the case we have calculated that $n=1.7\ 10^{25}$ m$^{-3}$ corresponding to the molar concentration $n=0.11\%$.

(b) For the defect center **$Fe^{3+}(Ta)-2V(O)$ (rhombic)** the parameters are: $T_c=29$ K, $\varepsilon(T_c)=400$, $\delta z=0.25$Å$=25\ 10^{-12}$ m, $Z=3$ elementary charges. For the case we have calculated that $n=0.6\ 10^{25}$ m$^{-3}$, corresponding to the molar concentration of about $0.04\%$.

The case **(b)** seem more preferable, since for the case $n<<1\%$.

In order to calculate ferroelectric and magnetic properties of $KTaO_3$ nanoparticles, one should know the temperature and size dependence of their dielectric permittivity. Since the latter determines the effective Bohr radius value, which in turn will determine the localized state radius.

The feature of the nanosized incipient ferroelectrics is the strong temperature and size dependence of their dielectric permittivity [44, 45]. In particular, the Barrett-type formula [46] was derived for the temperature dependence of the dielectric permittivity of incipient ferroelectrics nanospheres [47]:

$$\varepsilon(T,R) = \varepsilon_b + \frac{1}{\varepsilon_0} \left( \frac{\alpha_T \left( \frac{T_q}{2} \coth\left( \frac{T_q}{2T} \right) - T_0 \right) + \frac{4\sigma_S (Q_{11} + 2Q_{12})}{R} +}{\frac{\Lambda \varepsilon_0^{-1}}{\varepsilon_e R + \Lambda(\varepsilon_b + 2\varepsilon_e)} + \frac{g}{\left( \lambda + \sqrt{3g\varepsilon_0\varepsilon_b} \right)R} \left( 1 - \frac{3\Lambda \varepsilon_b}{\varepsilon_e R + \Lambda(\varepsilon_b + 2\varepsilon_e)} \right)} \right)^{-1} \quad (6)$$

Here the background permittivity $\varepsilon_b$=48, $R$ is the sphere radius, $T$ is the absolute temperature, $\varepsilon_0$ is the universal dielectric constant, $\alpha_T$ =2.02 $10^6$ m/(F K), Curie temperature $T_0$=13 K, Quantum temperature $T_q$=55 K [48, 49], electrostriction coefficients $Q_{11}$=0.087 m$^4$/C$^2$, $Q_{12}$= -0.023m$^4$/C$^2$, intrinsic surface stress (surface tension) coefficient $\sigma_S$=50 N/m [50], gradient coefficient g ~ $10^{-10}$ V·m$^3$/C, $\lambda$ is extrapolation length (typically unknown fitting parameter varied in the wide range from 0 to 100m). $\Lambda$ is either the screening length (~ 1-10 nm) or the free-bound charges separation distance (~ 0.1 nm) at the particle surface. Corresponding temperature dependences of the effective Bohr radius $a_B^*(T,R) = (1 + \varepsilon(T,R)) \frac{2\pi\varepsilon_0 \hbar^2}{|Z|\mu e^2}$ ($\mu$ is the fermion effective mass).

### B. Model of the defect driven magnetism in nanosized $KTaO_3$

Using the quantum-mechanical approach combined with the image charge method we calculated the lowest energy levels of the impurities and vacancies with two carriers (electrons or holes) localized in the point $\mathbf{r}_0 = (0,0,z_0)$ near the surface of KTaO$_3$ particle (Fig.10a). For the case of the cation vacancy the anion atom should be added to form the perfect host lattice, and its two electrons should be localized at the nearest anions. As a result a negatively charged defect (-2e) with two holes appears in the continuous media. The situation is vise versa for an anion vacancy: it can be modeled as a positively charged defect (+2e) with two electrons in the continuous media. We have assumed that oxygen vacancies are the source of the ferromagnetic ordering.

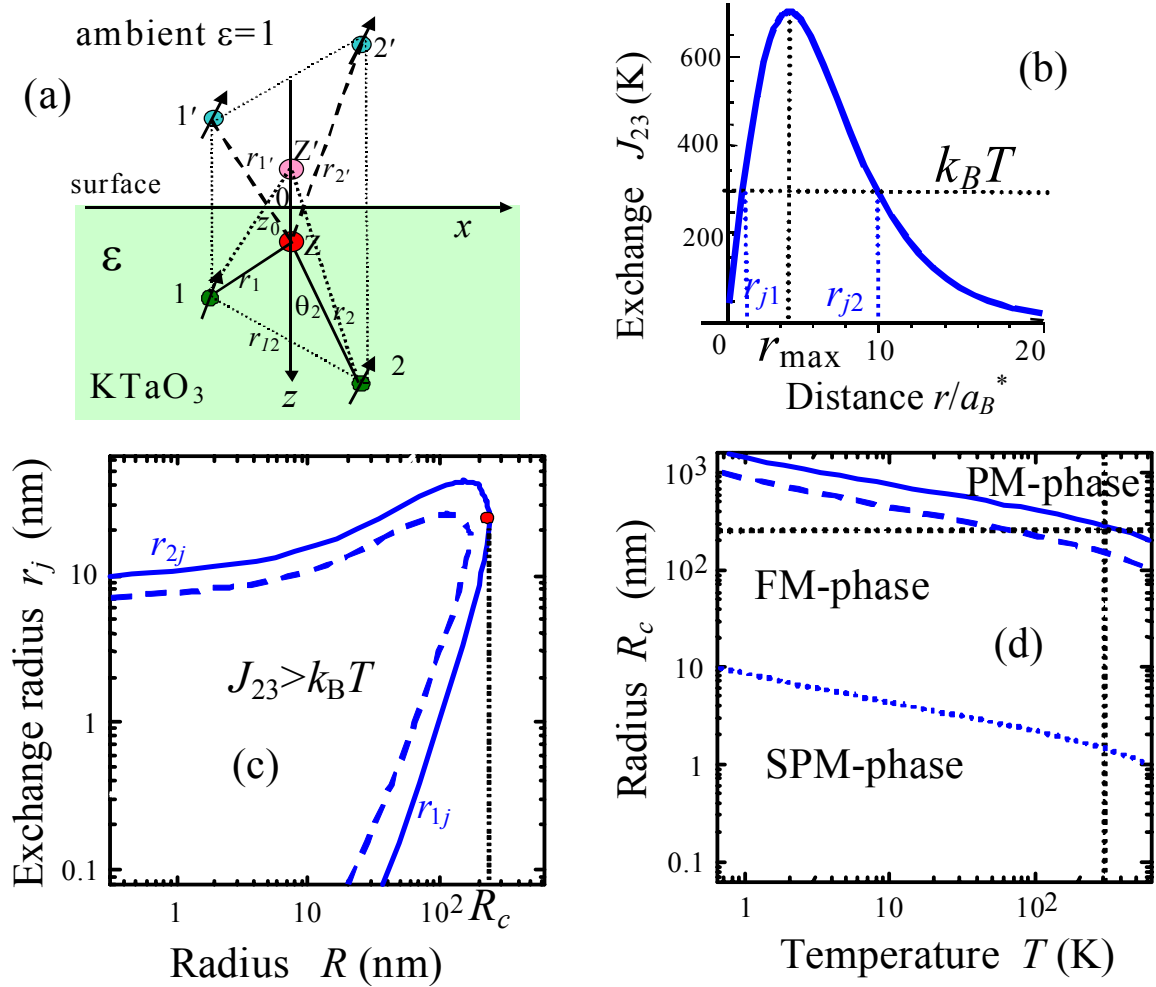

Fig.10. (Color online) (a) Defect at the distance $z_0$ under the KTaO$_3$ surface. Two carriers (electrons or holes) 1 and 2 (shown by green circles with arrows) are localized near the defect

with effective charge Ze (shown by red circle). Carrier image charges are shown as 1′ and 2′, defect image is Z′. In the continuum media approach impurity atoms, anion vacancies are considered as defects placed in a perfect host lattice. (b) Dependence of the exchange integral vs. the distance $r$ between the defects. (c) Dependences of the exchange radius $r_{j1}$ (dotted curves) and $r_{j2}$ (solid curves) on particle radius $R$ calculated for KTaO$_3$ nanoparticles at temperature $T$=300 K. (d) Phase diagram in coordinates particle radius – temperature calculated for KTaO$_3$ nanoparticles: FM – ferromagnetic, SPM – superparamagnetic, PM – paramagnetic phases. Parameters: effective mass $\mu=m_e$, $\sigma_S$=10 J/m$^2$, screening length $\Lambda$=10 nm (solid curves in plots c,d) and $\Lambda$=5 nm (dashed curves in plots c,d), gradient coefficient g=10$^{-10}$ V·m$^3$/C, background permittivity $\varepsilon_b$=48, ambient permittivity $\varepsilon_e$=1.

All calculations are based on the analytical results [51], obtained by the direct variational method used to solve the Schrödinger-Vanjie equation in the effective mass approximation allowing for the defect-carriers Coulomb interaction along with the image charges contribution near the surface. The approach, proposed much earlier [52], was used for the ground state calculations of the one-electron impurity center located near the flat surface. Note, that the image charges method is evolved in the continuum media approach, that requires the conception of the media dielectric permittivity $\varepsilon$ to describe the defect Coulomb potential, that is grounded when the characteristic size r$_d$ of the carrier localization at the defect center is several times larger than the lattice constant.[53, 54, 55] Since the defect is immovable the static dielectric permittivity $\varepsilon$ should be used.

Energy levels were calculated in the framework of the conventional perturbation theory as $E_{mn} = \left\langle \psi_{mn} \left| \hat{H} \right| \psi_{mn} \right\rangle$. The magnetic triplet state $E_{23}^T$ appeared the ground state of the anion vacancies (i.e. oxygen vacancies) in the vicinity of KTaO$_3$ surface, while the nonmagnetic singlet $E_{22}$ is the ground state under the distances $2 a_B^*$, where $a_B^*$ is the effective Bohr radius. For the measured values of the dielectric permittivity $\varepsilon(T,R) > 100$, the values $2a_B^* > 4$ nm even at $\mu = m_e$.

Note, that various defects (impurities and vacancies) concentration increases near the sample surface, in particular allowing for the strong lowering of their formation energies [56, 57, 58]. Therefore, the native vacancies should be present largely in the surface layer (more generally, in the vicinity of the surface), and the ferromagnetism observed in nanoparticles, made from KTaO$_3$ powder, could be strong enough due to the surface defects. For the considered case, the KTaO$_3$ nanoparticles contain the *shell* of a thickness $2a_B^*$ with magnetic oxygen vacancies.

Magnetic states of the cation defects (Fe, Ni ions) exist not only near the surface of the particle, but also in the deeper layers.

The defects can induce long-range ferromagnetism via the percolation nature [59]. The magnetic exchange radius $r_j$ was determined from the condition of the exchange energy $J_{23}(r_j)$ equality to the thermal energy at $J_{23}(r_j) = k_B T_r$, where $T_r$ (in Kelvins) is the actual temperature. The dependence of the exchange integral versus the distance $r$ between the defects is shown in Fig.10b. The exchange integral $J_{23}(r)$ is positive independently on the distance r. $J_{23}(r)$ moves to zero at $r \to 0$ (as calculated on p-states), has a pronounced maximum at distances $r = r_m$ and then vanishes exponentially with the distance increase $r >> r_m$. Exchange integral depends on the particle radius and temperature via the size and temperature dependence of dielectric susceptibility $\varepsilon(T,R)$. Namely, our numerical and analytical calculations showed:

$$J_{23}(r) = \frac{J_m}{(\varepsilon(T,R) + \varepsilon_e)^2} f\left(\frac{r}{(\varepsilon_e + \varepsilon(T,R))a_B}\right), \qquad (7)$$

where the amplitude $J_m$ is virtually independent on the dielectric permittivity at $\varepsilon(T,R) >> 1$ and $J_m/k_B \approx 858907$ K; $a_B = \frac{2\pi\varepsilon_0\hbar^2}{|Z|\mu e^2}$ is the "vacuum" Bohr radius. Positive dimensionless function $f(x)$ has the following Pade approximation $f(x) \approx \frac{x}{4} \exp\left(1 - \frac{x}{4}\right)$. The maximal value $f(r_m) = 1$, where $r_m \approx 4(\varepsilon_e + \varepsilon(T,R))a_B = 4a_B^*(T,R)$.

Note, that two roots ($r_{j1} < r_m < r_{j2}$) of the equation $J_{23}(r) = k_B T$ may exist at fixed temperature and particle radius $R < R_c(T)$. These roots tend to $r_m$ when the particle radius tends to the critical radius $R_c(T)$. Dependences of the exchange radius $r_j$ vs. the particle radius $R$ for KTaO$_3$ nanoparticles are shown in Fig.10c. It is seen from the Fig.10c that the exchange radius first increases with the particle radius increase (solid curves for $r_{j2}$), then reaches its maximum, and then decreases (dotted curves for $r_{j1}$) losing its sense at the critical radius $R_c$, once $J_{23}(r_m, R_c) < k_B T$.

Using Eq.(7), the critical radius $R_c(T)$ can be calculated from the condition $J(r_m) = \frac{J_m}{(\varepsilon(T,R_c) + \varepsilon_e)^2} = k_B T$. Since $\varepsilon(T,R) \gg 1$ for incipient ferroelectrics, the condition $J(r_m) = k_B T$ determines the ferromagnetic long-range order boundary $R_c(T)$ of the phase diagram in the coordinate's particle radius – temperature. Temperature dependence of the critical radius is shown in Fig.10d. It is seen from the Fig.10d that the radius decreases with the temperature increase, also its value depends on the screening length and surface tension.

Once the exchange radius is determined, and the particle radius is less than the critical radius $R_c(T)$, the percolation concentration $N^p$ of the random magnetic defects should be determined for a known radius $r_j$ as $N^p = 4/(3\pi r_j^3)$ if $r_j < 2a_B^*$ and $N^p = 4/(\pi r_j^2 a_B^*)$ if $r_j > 2a_B^*$. The concentration of the magnetic defects (oxygen vacancies) in KTaO$_3$ was estimated as $\sim 10^{25}$ m$^{-3}$ inside a magnetic shell with a thickness of 4 nm, corresponding to the molar concentration of about (0.06 – 0.1) %.

## V. DISCUSSION AND CONCLUSSION

Let us discuss the appearance of ferroelectricity in nanocrystalline potassium tantalate. The observed first-order Raman spectra and the softening of the lowest transverse $TO_1$ phonon at temperature less than a Curie temperature $T_c$ reveal the presence of polar nanoregions due to symmetry-breaking defects and a ferroelectric ordering at $T_c$=29 K. It was determined that the size of the nanoregions increases dramatically below $T_c$ and becomes about 10-14 lattice constants at 7 K. The molar concentration of symmetry-breaking defects was estimated to be 0.04-0.09%, depending on the phonon diffusion constant D~30-42 meV. Taking into account that the concentration of the unavoidable impurities of Fe and Ni are about 0.04 and 0.008 mole %, respectively, these atoms seem to be good candidates as the source of the polar nanoregions. A detailed analysis of the EPR spectrum showed that the paramagnetic $Fe^{3+}$ centers of axial and rhombic symmetry are presented in $KTaO_3$ nanopowder. However, the theoretical calculations showed, that only one type of center, namely a rhombic center, can induce a ferroelectric phase transition at $T_c$=29 K. For this case we have calculated that the molar concentration of the defects to be 0.04 %. Thus, we suggest that the polar nanoregions are induced by the Fe ions associated with oxygen vacancies, acting as the randomly distributed dipoles. Most likely, these are the centers of rhombic symmetry, in which $Fe^{3+}$ ion substitutes for $Ta^{3+}$ ion and forms an electric dipole of 0.25 Å length. The conclusion is consistent with the results of Sacle et al [40].

The ferromagnetism in $KTaO_3$ nanopowder is originated from both the magnetic state of the randomly distributed oxygen vacancies and the impurity defects (most likely Fe atoms). The concentration of the magnetic oxygen vacancies in $KTaO_3$ were estimated as 0.06-0.1%. The magnetic properties of nanoparticles are strongly size-dependent due to the size dependence of the incipient ferroelectric dielectric permittivity $\varepsilon(T, R)$ and the effective Bohr radius $a_B^*(T,R) \sim \varepsilon(T,R)$. The ferromagnetic long-range order is originated via the exchange coupling between the magnetic defects (vacancies and impurity atoms) for the particle radii $R$ less that the critical radius $R_c(T)$. It should be noted that undoped (or nominally pure) $KTaO_3$ single crystals are nonmagnetic and never exhibited a FMR spectrum. Only in the heavy (up to 3%) Fe-doped single crystals the FMR spectrum has appeared [37]. A polycrystalline $KTaO_3$ ceramic doped with a 0.25 mol. % of Fe showed the axial and cubic EPR spectra. And only a doping level of 1 mol. % of Fe resulted in the appearance of a FMR spectrum.[60] In the nanoscaled $KTaO_3$ the situation is different. Even in undoped nanopowder, so small concentration of the impurity Fe atoms as 0.04 mol. % makes an exchange coupling and a ferromagnetic ordering possible. The reason is that in the nanosized material the impurity ions as well as the oxygen vacancies are located mostly on the surface of the particle and thus form agglomerates with long-range magnetic ordering.

In conclusion, the defect driven ferroelectricity and magnetism are registered in nominally pure nanocrystalline potassium tantalate. Both ordering states are suggested to be due to the iron atoms and oxygen vacancies, which exist in the material as extrinsic and intrinsic defects. The concentration of defects increases near the particle surface and especially along the cube edges, accounting a cubic shape of the $KTaO_3$ nanoparticle. The results enable us to refer a nanosized $KTa(Fe)O_3$ to the class of multiferroics and assume that it could perform the magnetoelectric effect at T<29 K. To demonstrate the latter, additional measurements will be undertaken. On the other hand, first-principle calculations would be useful to construct the model of the surface magnetism in the nanosized $KTaO_3$. It was also established that the critical concentration of impurity defects necessary to provoke the appearance of the new phase states in the material strongly correlates with the size of the particle; as the size of the particle decreases, the critical concentration decreases as well. At this point we recommend to control carefully the incorporation of the "ubiquitous" iron in the host lattice especially in the nanotechnology processes.

## Acknowledgements

Authors are grateful to Prof. Glinchuk M.D. and Dr.Hab. Eliseev E.A. for useful discussions and fruitful ideas. Special thanks to Dr. Andriiko A.A. and Chernenko L. for synthesis of the nanocrystalline potassium tantalate and interest to the work.

---

* Author to whom correspondence should be addressed. Email address: golovina@isp.kiev.ua or isgolovina@yahoo.com